\documentstyle[12pt]{article}
\begin{document}

\newcommand{\ep}{\epsilon}
\newcommand{\be}{\begin{eqnarray}}
\newcommand{\ee}{\end{eqnarray}}
\newcommand{\bea}{\begin{eqnarray*}}
\newcommand{\eea}{\end{eqnarray*}}
\newcommand{\bh}{\beta \hbar}
\newcommand{\uv}{\mu\nu}

\rightline{{\bf CWRU-P15-96}}
\rightline{October 1996}
\baselineskip=16pt
\vskip 0.5in
\begin{center}
{\bf\large
DIRTY BLACK HOLES AND HAIRY BLACK HOLES}
\end{center}
\vskip0.2in
\begin{center}
Lawrence M. Krauss \footnote{also Department of Astronomy}, Hong Liu
\vskip .1in
{\small\it Department of Physics \\
 Case Western Reserve University\\
10900 Euclid Ave.,
Cleveland, OH 44106-7079}
\vskip .1in
and
\vskip .1in
Junseong Heo
\vskip .1in
{\small\it 
Physics Dept.\\ Yale University\\ New Haven CT 06511 }
\vskip 0.4in
\end{center}

\begin{abstract}

An approach based on considerations of the non-classical energy
momentum tensor outside the event horizon of a black hole provides additional
physical insight into the nature of discrete quantum hair on black holes and
its effect on black hole temperature. Our
analysis both extends previous work based on the Euclidean action techniques,
and corrects an omission in that work. We also raise several issues related to
the effects of instantons on black hole thermodynamics and the relation between
these effects and results in two dimensional quantum field theory.

\end{abstract}

\newpage
\baselineskip=21pt

\section{Introduction}

Semiclassical considerations of quantum fields in curved space backgrounds have
revolutionized our thinking about both classical and quantum
gravity.(i.e. \cite{hawking,preskill}) 
A great deal of work has been carried out in the past decade aimed at
addressing potential problems of untarity violation and information loss. The
jury is still out, but during this process many new insights have been gained
both about black hole physics, and about possible new Planck Scale phenomena, and
phenomena in higher dimensions.

One result which relies only on phenomena which
exist well below the Planck scale has been the recognition that black holes
can harbor ``quantum hair" (i.e.\cite{bowick,krauss,CPWa})---that is,
quantum mechanical observables can exist associated with black holes beyond
those allowed by the classical ``no hair theorems" \cite{bek,teit,adler}.

The prototype example of such quantum hair \cite{krauss} is quite simple. 
Consider an abelian U(1) gauge theory containing two matter fields $\eta$
and $\phi$ with charge $Ne$ and $e$ respectively.  If the field $\eta$
condenses at some energy scale $v$, then the gauge field will become
massive by the Higgs mechanism, and below this scale the effective theory
involving only the light field $\phi$ will have a residual discrete $Z_N$
symmetry.
This low energy broken symmetry theory also can contain stable strings
threaded by magnetic flux $2\pi /Ne$.  The scattering of $\phi$ quanta, with
charge $e$ from such strings is dominated by the Aharonov-Bohm effect
\cite{bohm,wilz} involving quantum phases uniquely determined by the product of
charge and flux, and thus allowing a determination of the total charge
modulo N which scatters off the string.

Since the quantum phases in question are global quantities, a $\phi$ quanta
which falls into a black hole will be measurable as such even after it falls
inside the event horizon of the black hole \cite{krauss,presk,wilz2}.  Such a
charged black hole therefore has ``discrete gauge quantum hair".
\cite{krauss,presk,CPWa}

Given the semiclassical relationship between entropy, area, and
temperature for black holes, one might expect that any
restriction on the number of microstates associated with a
given classical black hole macrostate, as would occur if one could measure
additional black hole quantum numbers, would have a concomitant effect on the
black hole's entropy, and hence its temperature.  In a beautiful series of
papers, Coleman, Preskill and Wilczek explored this possibility
\cite{ckpw,CPWa,CPWb}.  They uncovered an (exponentially small) effect on
temperature, but surprisingly the sign of the temperature change depended upon
the relative scales of the spontaneous symmetry breaking associated with the
quantum hair, and the inverse size of the black hole event horizon.  Moreover,
they also uncovered a new observable associated with quantum hair: a
non-classical electric field could exist and be measured outside of the
event horizon.

Motivated by the recognition that a non-classical field exists outside the
event horizon of a black hole endowed with discrete gauge hair, we
focus here on the energy momentum tensor outside the black hole. In this way,
one might hope to use Minkowski space arguments to get some additional insight
into the physics behind the effects of quantum hair which might not be manifest
in a Euclidean Action approach. 
  An appropriate Minkowski-space formalism was developed by Visser
\cite{viss} for treating ``dirty" black holes, where non-zero matter fields exist
outside the event horizon.  However, since the electric field associated with
discrete hair is non-classical, in the sense that it is not a solution of the
coupled vacuum Einstein-Maxwell equations, standard methods such as Visser's,
which require such solutions, cannot be applied
directly.

While the electric field generated outside the event horizon is not
a solution of the Minkowski field equations, the individual instantons whose
contributions sum to produce such a field are solutions of the coupled
Euclidean Einstein-Maxwell equations.  This suggests a hybrid approach, in
which we use the Visser formalism in Euclidean space, and then
focus on the effect of individual instantons, recognizing of course that
their contribution is negligible except to quantities which explicitly
depend on the discrete gauge charge, and which vanish in perturbation theory.

The Euclidean spacetime metric
generated by a static spherically symmetric  distribution of matter can be put
in the form:
\[
ds^{2} = e^{-2\phi(r)} ( 1 - \frac{b(r)}{r}) dt^{2} 
         + ( 1 - \frac{b(r)}{r})^{-1} dr^{2} + r^{2}d \Omega^{2}
\]
With the assumption  that the metric has an asymptotically 
flat geometry and an event horison, boundary conditions 
can be imposed as:
$\phi(\infty) = 0,  b(\infty) = 2GM_{BH},  b(r_{H}) = r_{H} $
where $M_{BH}$  is the mass of the  black hole and $r_{H}$ is the horizon
size.
Einstein's equations
can then be solved formally to give $b(r)$ and $\phi(r)$ in 
terms of the components of the energy momentum tensor.  Defining 
\[
T_{t}^{t} = \rho, \,\,\,\, T_{r}^{r} = \tau, \,\,\,\,
T_{\theta}^{\theta} = T_{\varphi}^{\varphi} = - \mu.
\]
the Hawking 
temperature and 
the horizon size of the black hole can be expressed as 
(i.e. see \cite{viss}):
\begin{eqnarray}
\frac{1}{\beta\hbar} & = & \frac{1}{4 \pi r_{H}}e^{- \phi (r_{H})} 
                     (1 - b'(r_{H})) 
\label{eq:HT} \\
r_{H} & = & 2 G M_{BH} + 8 \pi G \int_{r_{H}}^{\infty}dr \rho r^{2}
\label{eq:rh}
\end{eqnarray}

When the external matter contribution to the geometry is much 
smaller 
than that of the black hole, as will be the case of interest here, 
equations  (\ref{eq:HT}) and (\ref{eq:rh})
can be systematically expanded and the  lowest order
corrections to
 black hole thermodynamics 
can then be obtained.   
Define 
\begin{eqnarray*}
A  =  \frac{8 \pi G}{2} \int_{r_{H}}^{\infty} \frac{\rho - \tau}  
{r-r_{H}} r^{2} dr; \  
B  =  8 \pi G \rho_{H} r_{H}^{2}; \
m =  4 \pi \int_{r_{H}}^{\infty} (2\mu + \rho -\tau) r^{2} dr
\end{eqnarray*}
Expanding to first
order in 
$A$, $B$, and $(m/M_{BH})$ and using energy conservation one can derive an
expression for $\beta\hbar$ 
in terms of  only the components of the energy momentum
tensor and $M_{BH}$:
\be
\beta\hbar = 8\pi GM_{BH} ( 1 + \frac{m}{M_{BH}} - 2(A+B) + ...)
\label{eq:HTf}
\ee

Examining  (\ref{eq:HTf}) it is now clear that the sign of the correction to
the black hole temperature, for fixed mass, depends upon the relative sign of the
term 
$m/M_{BH} -2 (A+B)$.  It is precisely
this result which establishes the connection between the Weak Energy 
Condition (WEC), and the effect of fields outside the event horizon to the
black hole temperature.   All forms of classical matter which satisfy the
WEC also satisfy the relation $m/M_{BH} -2 (A+B) \ge 0$, implying that
classical matter outside the black hole can only lower the 
temperature\cite{viss}.
However, as we shall see, instanton contributions need not be of this
form.  Indeed, the non-classical electric field outside a black hole
endowed with discrete hair is precisely a manifestation of the fact
that quantum effects can violate the WEC.   

We now turn to the Euclidean Einstein-Abelian-Higgs system, the
prototypical example of quantum hair. This system
 has solutions corresponding to a
vortex sitting  in the 2-d Euclidean $r-t$ plane of a black hole. The two other
Euclidean dimensions $\theta , \phi$, (which would correspond to 
$z,t$ for a corresponding vortex in Minkowski space) are suppressed.  
As emphasized by by CPW, in
a Euclidean path integral formalism these
instanton solutions play a central role in
producing the observable non-classical effects of discrete charge outside
of the black hole event horizon, as the sum over these instantons includes
Aharonov-Bohm phases which are sensitive to the discrete charge contained 
the black hole. We wish to examine this effect in the
context of the formalism we have described above.
  
We use standard ansatz for these vortex solutions:
$ \phi = v f(r) e^{-i{2 \pi \over \bh}t}, 
A_{t} = {2 \pi \over \bh}{1 \over e}( 1 - a(r)), $
with boundary conditions:
$f(r_H) = 0, 
f( \infty ) = 1, 
a(r_H) = 1, 
a(\infty) = 0. \
$
$A_{t}$ satifies an equation which reflects the flux
quantization condition for vortices in the broken phase:
$
e \int_{0}^{\bh} dt \,\, A_{t}|_{r=\infty} = 2 \pi
$.
Following CPW one can consider two limiting cases, depending upon whether the
vortex width is much larger or smaller than the size of the event horizon.  The
virtue of equation \ref{eq:HTf} is that it lends itself
directly to such an analysis.  Competition among the
different terms as their
$r$-dependence varies, can lead, in different limits, to a different sign for
the correction to the black hole temperature. What actually occurs, however
depends subtlely yet crucially on the nature of the vortex solution in
curved space, as we shall demonstrate below.

The thin string limit is particularly simple to analyze in this context.  In
the thin string limit, the vortex width
$r_{s}
\ll r_{H}$, so that,
\bea
2(A+B) - \frac{m}{M_{BH}} & \approx & \frac{ 8\pi G}{r_{H}} \int_{r_{H}}^{\infty} (4\mu r r_{H} - 2\mu r^{2}) dr \\
& \approx & 16\pi G r_{H} \int_{r_{H}}^{\infty}  \mu dr \,
\eea
Then the correction to the Hawking temperature due
to the vortex instanton can directly be expressed as: 
$$
\bh \doteq 8\pi GM_{BH} [ 1 -
16\pi G r_{H} \int_{r_{H}}^{\infty}  \mu dr  ]\,
$$

In this limit, the vortex lies in the 
region $r \sim r_{H}$. One can then show (see \cite{kraussliu} for further
details) that 
$
A \sim B \sim 
{v^{2} \over M_{pl}^{2}}
$ 
and thus the correction to the Hawking temperature:
\bea
\bh 
& \sim &  8\pi GM_{BH} [ 1 - O({v^{2} \over M_{pl}^{2}})]
\eea
We see that the
effect of a single instanton in this limit is to raise the black
hole temperature.  Of course, we emphasize that to determine the thermal
effect of discrete charge one must sum over instantons, and thus go
beyond our formalism. (Note that if one does the summation\cite{CPWa}, 
the interference
between instantons and anti-instantons for the weighted action produces
an effect which is opposite in sign to that for the single instanton).
Nevertheless, the single instanton contribution to the temperature which
we calculate using the energy momentum formalism directly is identical with 
that determined by CPW in the thin string case based
on their estimates of the deficit angle and contributions to the action.

The thick string limit, in which $ r_s >> r_H$, is much more subtle,
precisely because in this limit the curvature associated with the 
sphere at the event horizon cannot be ignored, as in the thin string
limit. Put another way, we cannot accurately picture the instanton as
a vortex simply living in the two dimensions of a flat $ r-t$ plane.
If this persisted to be the case, one could use well known properties
of vortex solutions inside the core of the vortex, where the symmetry
is unbroken, along with the boundary conditions associated with the
magnetic flux carried in the core, to examine (\ref{eq:HTf}), and estimate
the instanton contribution.  However, if one does this, (as we confess we
first did), several anomalies arise.  In the first place, the lowest order
correction one finds to the black hole temperature is proportional to
$v^2/M_{pl}^2$, while CPW focus on a zeroth order contribution in this
limit.  In the second place, a straightforward application of this 
ansatz to a calculation of the action, which we always use as a check
of our approximation, yields a result which appears to be nonsensical---namely
that the instanton action is less than the Schwarzchild action.

The resolution of this paradox lies in consideration of the effects of
curved space, associated with the spherical surface at the event horizon.  In
this case, the vortex core behaves remarkably differently from the flat space
vortex (for further discussion see
\cite{kraussliu}).  The presence of extra $r^2$ contributions in the spherical
derivatives around the event horizon allow a vortex solution in which both the
gauge potential and magnetic field fall off exactly as in the unbroken theory
inside the core, so that the physics inside becomes largely insensitive to the
boundary conditions associated with the Higgs field behavior at the vortex
surface.  In other words, as $v \rightarrow 0$ the thick string limit
smoothly approaches the Reissner-Nordstrom case--there are no singular effects
due to boundary conditions at infinity in the curved space solution.
 
Without solving in detail for the vortex solution here, 
the net effect of the analysis is that no
spatial derivatives in solutions blow up at infinity, so that if we
define the quantities
\bea
y = {r \over r_{H}}, \,\,\,\, 
\ep^{2} = 2 e^{2} r_{H}^{2}v^{2}, \,\,\,\,
\beta_{0} = { \lambda \over 2e^{2}}
\eea
and explicity define the thick string limit by letting $\ep^{2}$
tend to zero, as the vortex becomes 
thicker and 
thicker with $\ep^{2}$ smaller
and smaller, the action reduces to that of 
Euclidean Reissner-Nordstrom black hole, and 
there are no small or large
parameters  other than $\ep^{2}$ in the energy momentum tensor.
We can thus expand these quantities in terms of $\ep^{2}$:
$ \rho  =  \rho_{0} + \ep^{2} \rho_{1} + \cdots, \ 
 \tau = \tau_{0} + \ep^{2} \tau_{1} + \cdots, \
 \mu = \mu_{0} + \ep^{2} \mu_{1} + \cdots $

One may wonder why one should bother to consider
corrections of order $\ep^{2}$  if the zeroth order
term in $\ep^{2}$ (the one which is calculated by CPW)
indeed gives the dominant contribution. The point is that 
this does not seem to be
guaranteed to be the case

The contributions from the 
zeroth order terms are identical to those in the 
case of the
Euclidean Reissner-Nordstrom black hole \cite{viss,CPWa} so that
\bea
\bh  \approx 
 \sim  8 \pi G M_{BH} 
( 1 + O({M_{pl}^{4} \over M_{BH}^{4}}) + \cdots )
\eea

The contributions from the first order terms yield:
\bea
\bh & = & 8 \pi G M_{BH} 
( 1 - { G \over 4 e^{2} r_{H}^{2}} \ep^{2} T  + \cdots) 
\eea
where T is a dimensionless quantity of $O(1)$ derived from (\ref{eq:HTf}) as
\bea
T=  \frac{ 4 e^{2} r_H}{8\pi}
 \int_{r_{H}}^{\infty} dr
[ (4\mu_{1} r r_{H} - 2\mu_{1} r^{2}) 
- (\rho_{1} - \tau_{1})r(r-r_{H})] 
\eea

Now we can compare the results from
the zeroth order terms and the first order terms.
The ratio of the first order (in $\ep^2$) contribution 
correction to the zeroth order 
(to both the Hawking Temperature and 
the action) is  
$$
\gamma = {\ep^{2}M_{BH}^{2} \over M_{pl}^{2}} =
 v^{2}M_{BH}^{4}/M_{pl}^{6}
$$ 

Now, recall that the thick string limit is
$$
\ep^{2} = v^{2}M_{BH}^{2}/M_{pl}^{4} \ll 1.
$$
However, for the semiclassical analysis of black
hole thermodynamics to be meaningful, $M_{pl}^{2}/M_{BH}^{2}$ 
has to be very small.  For sufficiently massive black
holes, it is certainly possible that both $\ep^2 \ll 1$ and $\gamma \gg 1$,
in which case the contributions
from the the first order terms cannot be neglected. 
For example, if we keep $\ep^2$ fixed but
let $(M_{BH}/M_{pl})^{2} \rightarrow \infty $ (which requires also
making $v^2 \rightarrow 0$ ), then $\gamma \rightarrow \infty$, so that while
both the zeroth and first order contributions go to zero, the first
order piece becomes arbitrarily large compared to the second. Stated
another way, the limit in which only the Reissner-Nordstrom 
piece is considered, as was done by CPW, is { \it not}
 the generic thick string
limit, but is rather the limit $\gamma \ll 1$. 

It is perhaps surprising that for sufficiently large black holes
the first order terms in $\ep^2$ may become comparable or larger than the
zeroth order terms.  However recalling our heuristic discussion earlier
in this section, this effect is perhaps understandable.  For larger black
holes the curvature at the event horizon becomes progressively
smaller.  While it may be true that the $v \rightarrow 0$ limit goes
smoothly to the Reissner-Nordstrom case, increasing
the black hole mass reduces the curvature effects at the horizon which
are responsible for the domination of the Reissner-Nordstrom
contribution compared to vortex symmetry breaking contribution
proportional to the vev of the Higgs field---namely, 
$\ep^2$ must be correspondingly reduced as the black hole mass increases
in order for the first order contribution in $\ep^2$ to be negligible.

Finally then, we may ask what the sign of the first order term in the
expression for the black hole temperature given above is.  The 
sign of $T$ is in general indeterminate.  However,
if we make the anzatz that the first order term takes a form similar
to that which would occur for the flat space vortex, one can
show that this quantitity is manifestly
negative, and hence the contribution of this piece to the Hawking temperature
would be of the same sign as the zeroth order contribution----namely instantons
in the thick string limit generically cool down a black hole, the opposite of
the thin string result.  Thus the general observation of CPW on the nature of
the effect is correct, even if the overall order of the dominant contribution
may not be what they calculated.   Note, also that in all cases the dominant
contribution to the action is positive, so the semiclassical instanton
approximation is stable.

Our
results indicate that one may fruitfully extend Minkowski space methods
designed to probe the effects of classical fields outside the event horizon on
the thermodynamical properties of black holes to the Euclidean regime of
semiclassical phenomena.  This allows a more
intuitive physical picture of the origin of such effects.  It may be
useful in exploring the nature of other semiclassical contributions to black
hole thermodynamics beyond those considered here associated with quantum hair.

We conclude with a remark which is more relevant to the Euclidean formalism
directly, and to the instanton sum which results in a non-classical electric
field outside the event horizon.  Such a non-classical electric field induced by
instanton effects is not unique to black holes.  Indeed, the prototypical
example occurs in a two dimensional Abelian Higgs model.   In this case, in the
presence of a topological term, Euclidean instantons induce a non-zero
non-classical background electric field (i.e. \cite{raj}).  Note that there are no
corresponding Euclidean instantons in the four dimensional Abelian Higgs model
in flat space.  However, the presence of a black hole event horizon alters the
topology of the corresponding Euclidean continuation so that instantons of the
type examined here both exist, and, if the black hole is charged, can produce
observable effects.  We believe the analogy between the two dimensional Abelian
Higgs model with a topological term and a black hole with discrete hair can be
made exact, and are currently exploring this issue \cite{kraussliu}.  If this is the
case, discrete hair may be cast in a different light, which may bear a closer
relation to other results associated with two dimensional field theories. 

\bigskip
We would like to thank Sidney
Coleman, John Preskill, Tanmay Vachaspati and  Frank Wilczek for discussions and
essential insights.  LMK and HL are supported by the DOE and funds from Case
Western Reserve University.

\end{document}